\documentclass[
preprintnumbers,a4paper,superscriptaddress,aps,showpacs]{revtex4}
\usepackage{graphicx} 
\begin{document}
\def\la{\langle}
\def\ra{\rangle}
\def\om{\omega}
\def\Om{\Omega}
\def\vep{\varepsilon}
\def\wh{\widehat}
\def\da{\dagger}
\def\tr{\rm{Tr}}
\newcommand{\beq}{\begin{equation}}
\newcommand{\eeq}{\end{equation}}
\newcommand{\beqa}{\begin{eqnarray}}
\newcommand{\eeqa}{\end{eqnarray}}
\newcommand{\intf}{\int_{-\infty}^\infty}
\newcommand{\into}{\int_0^\infty}
\preprint{EHU-FT/0106}
\title{Asymptotic behaviour of the probability density 
in one dimension} 
\author{J. A. Damborenea}
\affiliation{Fisika Teorikoaren Saila,
Euskal Herriko Unibertsitatea,
644 P.K., 48080 Bilbao, Spain}
\affiliation{Departamento de Qu\'\i mica-F\'\i sica,
Universidad del Pa\'\i s Vasco, Apdo. 644, Bilbao, Spain} 
\author{I. L. Egusquiza}
\affiliation{Fisika Teorikoaren Saila,
Euskal Herriko Unibertsitatea,
644 P.K., 48080 Bilbao, Spain}
\author{J. G. Muga\footnote{
Email: qfpmufrj@lg.ehu.es}}
\affiliation{Departamento de Qu\'\i mica-F\'\i sica,
Universidad del Pa\'\i s Vasco, Apdo. 644, Bilbao, Spain} 
%

\begin{abstract}
We demonstrate that the probability density of a quantum state moving
freely in one dimension may decay faster than $t^{-1}$. $t^{-2}$ and
$t^{-3}$ dependences are illustrated with analytically solvable
examples.  Decays faster than $t^{-1}$ allow the existence of dwell
times and delay times.\\
\end{abstract}

\maketitle

The study of the asymptotic, long-time behaviour of wave packets, and
related quantities such as decay and survival probabilities has a long
history. Recently Unnikrishnan$^{\ref{Unnikrishnan98},\ref{Unnikrishnan97}}$
has 
proposed several examples of decay of the wavefunction which are
slower than the $t^{-1/2}$ decay that follows from straightforward
dimensional analysis. In particular, Unnikrishnan has put forward
behaviours of the type $t^{-1/2+\mu}, 0\le\mu<1/4$ and $t^{-1/2}\ln
t$. In this paper we complement this study by giving examples of
decays which are faster than those considered in
Refs.\ref{Unnikrishnan98},\ref{Unnikrishnan97}.
We give a procedure for obtaining
systematically different kinds of faster decays, which we illustrate
with two explicit examples, and then discuss the failure of the
arguments that suggest that only the $t^{-1/2}$  
(or slower) decay  can take place.

An important consequence of the existence of faster decays is the possibility 
to define {\it dwell times},$^{\ref{Buttiker83},\ref{Nussenzveig00}}$ 
\beq\label{tauq}
\tau_D(a,b)\equiv\intf dt \int_a^b dx\,|\psi(x,t)|^2,
\eeq
and {\it delay times} for one dimensional motion, the later being
defined as the difference between the dwell time with and without an
interaction potential.$^{\ref{Smith60}}$  

The classical counterpart of (\ref{tauq}) is$^{\ref{MBS92}}$  
\beq\label{dwelltcl}
\tau_D(a,b)_{classical}=\int_{-\infty}^{\infty}dt\int_a^b dx\,
\varrho(x,t), 
\eeq
where $\varrho(x,t)$ is the probability density of an ensemble of
independent particles. $\tau_D(a,b)_{classical}$ is the average over
the ensemble of the time that each particle trajectory spends between
$a$ and $b$.  In other words, this is an average ``dwell'' or
``sojourn'' time in the selected region. Even though the
interpretation of (\ref{tauq}) is not as straightforward, since the
individual member of the quantum ensemble is not associated with a
trajectory in the standard interpretation, the quantum dwell time
is in any case
recognized as an important characteristic quantity of the wave
packet.$^{\ref{MT95}}$
Similarly, the delay time is, together with the
transmittance, one of the basic quantities that summarize the
essential aspects of wave packet scattering, also in the stationary
limit.  $\tau_D$ can be written in several ways, in particular as
\beq
\tau_D(a,b)=\intf dt\,P_{ab}(t)
=\la \psi(t=0)|\wh{T}_D|\psi(t=0)\ra,
\label{1}
\eeq
where  $P_{ab}(t)=\int_a^b dx\,|\psi(x,t)|^2$,
$\wh{T}_D$ is the {\it sojourn time operator},  
\beq
\wh{T}_D=\intf dt\, e^{i\wh{H}t/\hbar}\wh{D}(a,b)e^{-i\wh{H}t/\hbar},
\label{soj}
\eeq
$\wh{H}$ is the Hamiltonian, and
$\wh{D}(a,b)$ is the projector onto the selected space region,
\beq
\wh{D}(a,b)=\int_a^b dx\,|x\ra \la x|.
\eeq
Unless $P_{ab}(t)$ decays faster than ${t}^{-1}$, the dwell time will
diverge.  The existence of a potential function leads generically to
an asymptotic decay $\sim {t}^{-3}$, as discussed in Ref. \ref{MDS95} (see
also Ref. \ref{Amrein01} and references therein).  The situation is
however rather different for the case of free motion, with Hamiltonian
$\wh{H}_0$, for which the sojourn time operator (\ref{soj}) takes 
in momentum representation the
form 
\beq
\wh{T}_{D,H_0}=\sum_{\alpha=\pm}\int_{-\infty}^\infty dp\, \frac{mh}{|p|} 
|p\ra\la p|\wh{D}|\alpha p\ra\la \alpha p|.
\eeq
Note the presence of a $|p|^{-1}$ factor that will make the dwell time 
divergent unless the state vanishes at $p=0$.  
The fact that generically the dwell time for free motion will diverge
also follows from the dependence on $t^{1/2}$ of the corresponding
propagator,
\begin{equation}
\langle x|e^{-i\wh{H}_0t/\hbar}|x'\rangle=
\left(\frac{m}{iht}\right)^{1/2}e^{im(x-x')^2/2\hbar t}, 
\label{fmp}
\end{equation}
since this leads to a $t^{-1}$ dependence of the probability density 
$|\psi(x,t)|^2$ in a generic case.

There exists an equivalent scaling argument to understand this fact:
in the generic free motion case there are no dimensional magnitudes
other than $\hbar$ and $m$ present, and therefore no relevant
magnitude with dimensions of time can exist. Since
$P_{ab}(t)$ is, dimensionally speaking, an inverse time, it has to
scale with $t^{-1}$ in the absence of such a quantity. In the case of
interaction, the interaction length $l_0$ comes into play, thus
modifying this generic scaling, since the magnitude $m l_0^2/\hbar$ is
a characteristic interaction time.

On the other hand, if the initial data is such that a characteristic
length or a characteristic momentum exists, and it cannot be
eliminated by translations or Galilean boosts, it is likely that the
large time asymptotics will be modified. An alternative way of looking
into this possibility is by noticing that, using Eq. (\ref{fmp}), 
an integral over $x'$ is
required to obtain $\psi(x,t)$, which entails that the asymptotic
$t^{-1/2}$ dependence of the wavefunctions is not guaranteed in all
cases: the limit $t\to\infty$ and the integral do not necessarily
commute.  In particular, Unnikrishnan has found examples of slower
decay,$^{\ref{Unnikrishnan97},\ref{Unnikrishnan98}}$
and we shall justify and
illustrate decays faster than $t^{-1}$ when the momentum amplitude
$\la p|\psi\ra$ vanishes at $p=0$.

The large time asymptotics of $\psi(x,t)$ will depend on the behavior
near the ``critical point''$^{\ref{BH86}}$ $p=0$,
\beq\label{basic}
\psi(x,t)=\la x|\psi(t)\ra=\intf dp\, 
\la x|p\ra e^{-ip^2t/2m\hbar} \la p|\psi(0)\ra, 
\eeq
whereas larger momenta are responsible of the short time behaviour, as
happens when considering a classical ensemble too.  To determine the
large time behavior we shall assume that $\la p|\psi(0)\ra$ may be
expanded around the origin (this is not the case for the ``Cauchy
state'' chosen by Unnikrishnan to show slow decay$^{\ref{Unnikrishnan97},
\ref{Unnikrishnan98}}$ - in fact, this state presents a
logaritmic singularity at the origin of momenta).  Let us assume in
particular that the state belongs to the subspace of positive momenta,
and that it vanishes at $p=0$. Without loss of generality we set
$x=0$,

\beq 
\psi(0,t)\sim
h^{-1/2}\int_0^\infty dp\,e^{-ip^2t/2m\hbar} [ap+bp^2+...]. \label{general}
\eeq
Now make the change of variables  
\beq\label{uf}
u\equiv p/f,\,\,\,\,\,f\equiv(1-i)\sqrt{(m\hbar/t)}.
\eeq
On deforming the contour to the steepest descent path in the fourth quadrant
(where $u$ is real), we obtain 
\beq
\la x=0|\psi(t)\ra\sim h^{-1/2}f
\int_0^\infty du\, e^{-u^2}
[afu+b(fu)^2+...],
\eeq
which leads to a large time dependence $|\psi(x=0,t)|^2=O[t^{-2}]$ for
$a\neq 0$, or $O[t^{-3}]$ for $a=0, b\neq 0$.

\begin{figure}[tbp]
\includegraphics[height=8cm]{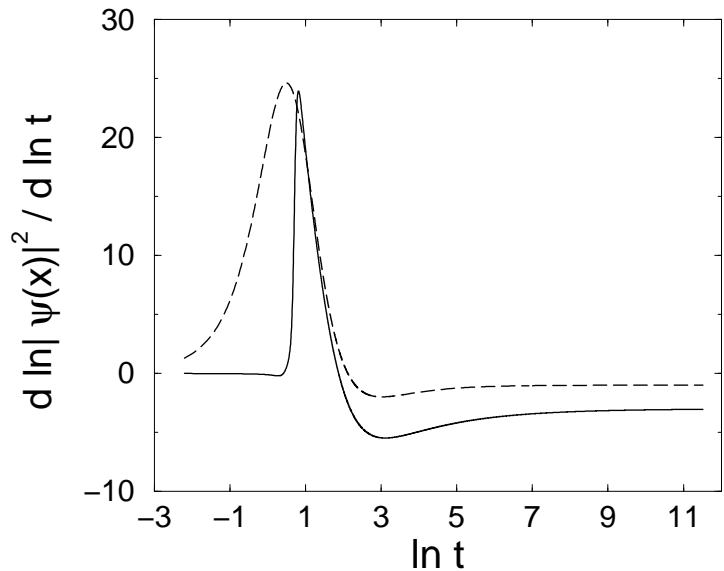}
\caption{\label{asym}
$d \ln |\la x |\psi(t)\ra|^2/d \ln t$ versus $d\ln t$ 
for two different wave packets: one of them is given by (\ref{pe0}),
({\it solid line}), and the other one is a 
a Gaussian wave packet,   
$\la p|\psi(0)\ra=C'e^{-\delta^2(p-p_0)^2/\hbar^2
-ipx_0/\hbar}$ ({\it dashed line}), where $C'$ is the normalization 
constant. The parameters are     
$p_0=1$, $x_0=-10$, $\alpha=0.5$, $\delta=1$, $x=0$, $m=1$ 
(all quantities in atomic units).
Note the asymptotic dependences of the probability densities:   
$t^{-3}$ and $t^{-1}$ respectively.}
\end{figure}

An example of the later case is given by the state$^{\ref{MPL99}}$ 
\beq\label{pe0}
\la p|\psi(0)\ra
=C(1-e^{-\alpha p^2/\hbar^2})e^{-\delta^2(p-p_0)^2/\hbar^2-ipx_0/\hbar} 
\Theta(p),
\eeq
for which $a=0$, and $b\ne 0$. ($C$ is the normalization constant and 
$\Theta$ is the Heaviside ``function''.)    
The time dependence may be obtained explicitly from (\ref{basic}), 
\beq 
\la x|\psi(t)\ra=
\frac{Ch^{1/2}}{4\pi^{1/2}e^{k_0^2\delta^2}}
\bigg\{\frac{w(-ig/2A^{1/2})}{A^{1/2}}
-\frac{w[-ig/2(A+\alpha)^{1/2}]}{(A+\alpha)^{1/2}}\bigg\}.
\eeq
Here $w(z)=e^{-z^2}{\rm erfc}(-iz)$, $k_0=p_0/\hbar$,
$A=\delta^2+i\hbar t/(2m)$, and $g=i(x-x_0)+2k_0\delta^2$.  From the
asymptotic properties of the $w$ function, see the appendix, it
follows that $|\psi(x,t)|^2=O[t^{-3}]$ as illustrated in Figure 1,
where we have also depicted the $t^{-1}$ decay corresponding to a
minimum-uncertainty product state.

The intermediate decay $t^{-2}$ in the probability density is obtained
for example for the initial wavefunction
\[
\psi(x,0)={{N}\over{\left(x +i \alpha\right)^2}}\,,
\]
where $\alpha$ is real and positive, and $N=\sqrt{2\alpha^3/\pi}$ is
the normalization factor. As a matter of fact, this is a direct
counterexample to the statement of Unnikrishnan that spatial decay
faster than $|x|^{-1}$ necessarily entails temporal decay of the form
$t^{-1/2}$ for the wavefunction.  The Fourier transform of this
initial wave packet is
\[
\psi(p,0)=-2\left(\frac\alpha\hbar\right)^{3/2}\Theta(p)
p  e^{-\alpha p/\hbar} \,,
\]
from which it is straightforward to conclude the decaying behaviour
mentioned above. The explicit expression for all times can be computed
in terms of ${}_1F_2$ hypergeometric functions.

Another example with $t^{-1}$ decay in the wavefunction is given by
considering the following initial wavefunction in momentum
representation:
\[\psi(p,0)=\frac2{\pi^{1/4}\beta^{3/4}}
\Theta(p) p e^{-p^2/2\beta}\,.
\]
The solution of Schr\"odinger's equation with this initial condition,
now written in coordinate space, is
\[\psi(x,t)=
\frac{m \beta^{1/4}2^{1/2}\hbar^{1/2}}{\pi^{3/4}
(i\beta t + m\hbar)} \left[1+i\pi^{1/2}\xi w(\xi)\right]\,,
\]
where 
\[
\xi=\frac{x}{2\hbar\left(\frac1{2\beta}+\frac{it}{2m\hbar}\right)}\,.
\]
Again using the series expansion of the $w$ function, one can easily
check the asymptotic behaviour mentioned above. In fact, in this
particular example it is easy to derive an explicit expression for $d
\ln |\la x |\psi(0,t)\ra|^2/d \ln t$, which results in
$-2/(1+m^2\hbar^2/\beta^2 t^2)$. 

In conclusion, we have shown that the asymptotic, large time 
decay laws of the free-motion one-dimensional wavefunction 
described by Unnikrishnan$^{\ref{Unnikrishnan98},\ref{Unnikrishnan97}}$
(as  $t^{-1/2}$ or slower) are not exhaustive. Faster decays
($t^{-1}$ and $t^{-3/2}$) have been demonstrated with analytically solvable 
examples and may be expected for wave packets that vanish 
at zero momentum.  
These faster decays lead to finite dwell times and allow to 
give a meaning to the ``lifetime'' (delay time) matrix of
Smith$^{\ref{Smith60}}$  
in one dimension in terms of the difference of dwell times with and without 
interaction potential.$^{\ref{Muga02}}$

\begin{acknowledgments}
This work has been supported by Ministerio de Ciencia y
Tecnolog\'{\i}a (Grants BFM2000-0816-C03-03 and AEN99-0315), The
University of the Basque Country (Grant UPV 063.310-EB187/98), and the
Basque Government (PI-1999-28). J.A.D. acknowledges a studentship of
The Basque Government.
\end{acknowledgments}

\appendix*
\section{Properties of $\lowercase{w}$-functions}
The $w-$function is an entire function defined in terms of
the complementary error function as$^{\ref{AS72}}$
$w(z)=e^{-z^2}{\rm erfc}(-iz)$.  
Its integral expression is 
\beq\label{wint}
w(z)=\frac{1}{i\pi}\int_{\Gamma_-}\frac{e^{-u^2}}{u-z}du,
\eeq
where $\Gamma_-$ goes from $-\infty$ to $\infty$ passing below the
pole at $z$.  From (\ref{wint}) two important properties are deduced,
\beqa\label{A3}
w(-z)&=&2e^{-z^2}-w(z)\\
\noalign{\hbox{\rm and}}
w(z^*)&=&[w(-z)]^*.
\eeqa
The derivative of the $w$-function can be expressed in terms of the
$w$-function itself,

\begin{equation}\label{A5}
w'(z)=-2zw(z)+\frac{2i}{\sqrt{\pi}}.
\end{equation}
Finally, $w(z)$ has the series expansion
\begin{equation}\label{wser}
w(z)=\sum_0^\infty\frac{(iz)^n}{\Gamma(\frac{n}{2}+1)}.
\end{equation}

\begin{enumerate}
\item\label{Unnikrishnan98}
{K.}~{Unnikrishnan}, ``An exhaustive analysis
of the asymptotic time dependence of wave packets in one dimension'',
  Am. J. Phys. \textbf{66},
  {632-633} (1998).

\item\label{Unnikrishnan97}
K.~{Unnikrishnan}, ``On the
asymptotic decay for wave packets in free space'',
{Am. J. Phys.} \textbf{65},
{526-527} (1997).

\item\label{Buttiker83}
{M.}~{B{\"{u}}ttiker}, ``Larmor precession and the traversal time for 
tunneling'', 
Phys. Rev. B \textbf{{27}},
{6178-6188} {1983}.

\item\label{Nussenzveig00}
{H.~M.} {Nussenzveig}, ``Average dwell time and tunneling''
{Phys.Rev. A} \textbf{{62}},
{042107} ({2000}).

\item\label{Smith60}
{F.~T.} {Smith}, ``Lifetime matrix in collision theory''
{Phys. Rev.} \textbf{118},
 {349-356} (1960).

\item\label{MBS92}
{J.~G.} {Muga},
  {S.}~{Brouard}, {and}
  R.~{Sala}, ``Transmission and reflection tunneling times''
  {Phys. Lett. A} \textbf{{167}},
  {24-28} ({1992}).

\item\label{MT95}
{H.}~{Mizuta} {and}
  {T.}{Tanoue},
  {The Physics and Applications of Resonant Tunnelling
  Diodes} (Cambridge University Press,
  {1995}).

\item\label{MDS95}
J.~G. {Muga}, 
  {V.} {Delgado}, {and}
  {R.~F.} {Snider}, ``Dwell time and asymptotic behavior of the 
  probability density''
  {Phys. Rev.} \textbf{52},
  {16381-16384} ({1995}).

\item\label{Amrein01}
{W.~O.} {Amrein}, ``Controversy about time decay''
  (2001), {quant-ph/0104049}.

\item\label{BH86}
{N.}~{Bleistein} {and}
{R.~A.} {Handelsmann},
{Asymptotic expansions of integrals}
  ({Dover}, {New York}, {1986}), p. 84.

\item\label{MPL99}
{J.~G.} {Muga},
 {J.~P.} {Palao}, {and}
 {C.~R.} {Leavens}, ``Arrival time distributions and perfect absorption
 in classical and quantum mechanics'',
 {Phys. Lett. A} \textbf{253},
  {21-27} (1999), quant-ph/9803087.

\item\label{AS72}
{M.}~{Abramowitz} {and}
  {I.}~{Stegun},
  {Handbook of Mathematical Functions}
  (Dover, New York, 1972), p. 297. 

\item\label{Muga02}
J. G. Muga, ``Characteristic times in scattering theory'', in 
``Time in Quantum Mechanics'', edited by 
J. G. Muga, R. Sala Mayato and I. L. Egusquiza
(Springer-Verlag, Berlin, 2002), Chap. 2, to appear.  
\end{enumerate}


\end{document}